\RequirePackage{fix-cm}

\documentclass[twocolumn,epjc3]{svjour3}  
\smartqed  
\RequirePackage{graphicx}
\RequirePackage{cite}

\RequirePackage{hyperref}
\RequirePackage{color,xcolor}
\definecolor{nicered}{rgb}{0.5,0.,0.}
\definecolor{nicegreen}{rgb}{0.,0.5,0.}
\definecolor{niceblue}{rgb}{0.,0.,0.5}
\hypersetup{
	colorlinks=true,
	linkcolor=black,
	filecolor=nicegreen,      
	urlcolor=niceblue,
	citecolor=niceblue,
}

\journalname{Eur. Phys. J. C}
\begin{document}

\title{Double and multiple bangs at tau neutrino telescopes
}
\subtitle{A novel probe of sphalerons with cosmogenic neutrinos}


\author{Guo-yuan Huang\thanksref{e1,addr1} 
}

\thankstext{e1}{e-mail: guoyuan.huang@mpi-hd.mpg.de}


\institute{Max-Planck-Institut f{\"u}r Kernphysik, Saupfercheckweg 1, 69117 Heidelberg, Germany \label{addr1}
}

\date{Received: date / Accepted: date}

\maketitle

\begin{abstract}
In light of the exciting campaign of cosmogenic neutrino detection, we investigate the double and multiple tau bangs detectable at future tau neutrino telescopes. 
Such events are expected from the Standard Model (SM) higher-order processes, which can be easily identified  with broad techniques anticipated at future tau neutrino telescopes.
We find that SM perturbative processes can already contribute  observable double-bang events to telescopes with a sensitivity of collecting $\mathcal{O}(100)$ cosmogenic neutrino events.
The detectable but suppressed rate in fact makes the double and multiple bangs an excellent probe of SM unknowns and possible new physics beyond.
As a case study, the nonperturbative sphaleron process, which can copiously produce multiple tau bangs, is explored. 
\end{abstract}

\section{Introduction}
The energy frontier of particle collisions has been pushed with continuous efforts.
Other than artificial colliders, the free cosmic particle fluxes have offered us another opportunity to study particle physics.
Remarkably, many cosmic rays and  gamma rays at ultrahigh energies have been observed with the ground-based arrays~\cite{PierreAuger:2020qqz,TelescopeArray:2021zox,Ivanov:2021mkn,LHAASO:2021cbz,LHAASO:2021crt,LHAASO_nature}.
Ultrahigh energy neutrinos, as the most elusive messenger associated with this frontier, were also detected up to $\mathcal{O}(\rm PeV)$ energies by the IceCube Observatory~\cite{IceCube:2013cdw,IceCube:2020acn,IceCube:2020wum}.
Yet, the cosmogenic neutrinos with energies as high as ${\rm EeV}\equiv 10^3~{\rm PeV}$, which are guaranteed by the Greisen-Zatsepin-Kuzmin mechanism~\cite{Greisen:1966jv,Zatsepin:1966jv,Beresinsky:1969qj}, have not been observed~\cite{IceCube:2018fhm,PierreAuger:2019ens,ANITA:2019wyx}.
%


In the near future, robust observations of the EeV neutrino flux are highly anticipated with a large number of experimental programs~\cite{Huang:2021mki,Ackermann:2022rqc,Abraham:2022jse,Neronov:2016zou,GRAND:2018iaj,Otte:2018uxj,Otte:2019aaf,ARA:2019wcf,Abarr:2020bjd,Anker:2020lre,RNO-G:2020rmc,IceCube-Gen2:2020qha,POEMMA:2020ykm,Romero-Wolf:2020pzh,Wissel:2020fav,Wissel:2020sec,Hallmann:2021kqk,Ogawa:2021dK,deVries:2021BA}; see Ref.~\cite{Huang:2021mki} for a recent compilation.
Among them, a promising experimental class is the tau neutrino telescope~\cite{Berezinsky:1975zz,Domokos:1997ve,Domokos:1998hz,Capelle:1998zz,Fargion:1999se,Fargion:2000iz,LetessierSelvon:2000kk,Feng:2001ue,Kusenko:2001gj,Bertou:2001vm,Cao:2004sd,Baret:2011zz}, which detects the decay products of tau leptons converted from tau neutrinos.
With this new energy frontier being established, an unprecedented center-of-mass (COM) energy of more than $43~{\rm TeV}$  for the neutrino-nucleon collision can be reached, much higher than that has been achieved in laboratories. 
This guarantees a new place to probe the Standard Model (SM) of particle physics and possible new physics beyond.
However, general issues of its particle physics potential are the large SM background and systematic uncertainties of the initial neutrino flux~\cite{Jezo:2014kla,Huang:2019hgs,Denton:2020jft,Soto:2021vdc,Huang:2021mki,Valera:2022ylt,Huang:2022pce,Esteban:2022uuw}. In such a case, new physics effects mostly manifest themselves indirectly by affecting the energy and angular distributions of tau events.

In this work, we explore a powerful topology at tau neutrino telescopes, the double- and multi-bang ($\geq 3$) events. 
Such events can be generated from higher-order (perturbative) processes in the SM, which by themselves are observable with the upcoming experimental programs. Furthermore, the relatively suppressed rate makes such events a distinctive probe of various physics unknowns. 
As a concrete example, we study the production of the well-motivated sphaleron, expected from SM nonperturbative processes, by the neutrino-nucleon scattering with an EeV incoming neutrino flux.


\begin{figure}[t!]
	\begin{center}
		\includegraphics[width=0.48\textwidth]{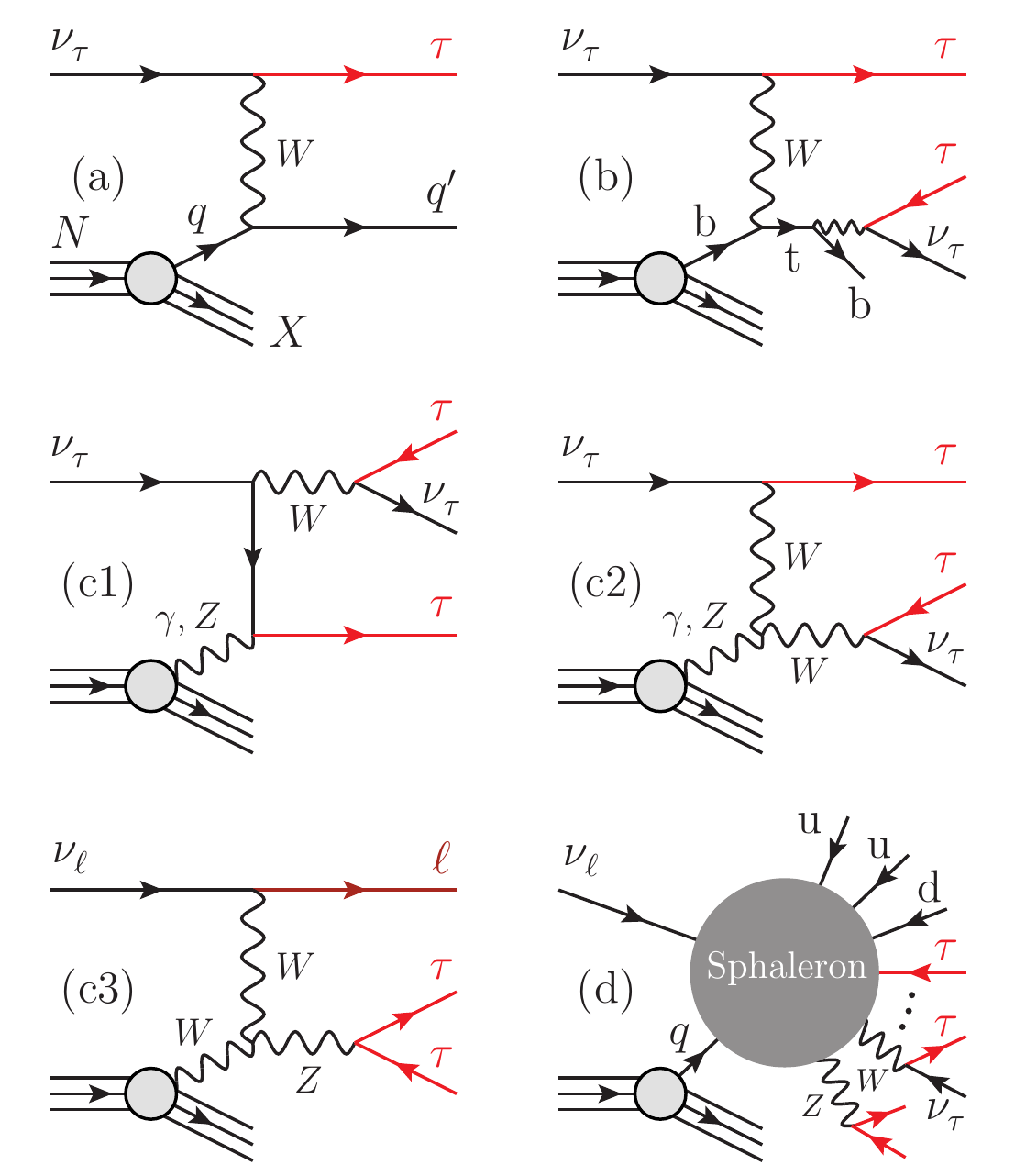}
	\end{center}
	\vspace{-0.3cm}
	\caption{Representative Feynman diagrams for the neutrino-nucleus scattering, with the productions of a single tau at the leading order (a) as well as double (b, c1 and c2) and triple taus (c3). The sphaleron process (d) producing multiple taus is also illustrated. Note that not all diagrams with the same final state are shown. }
	\label{fig:feyn}
\end{figure}

\section{Distinctive double and multiple tau bangs}
Tau neutrino telescopes are designed to observe the extensive air shower produced by the decay of a tau lepton (a ``single bang''),  originated from the charged-current (CC) deep-inelastic scattering (DIS) $\nu^{}_{\tau} + N \to \tau + X$, with $N$ being the nucleus and $X$ being the nuclear remnant.
The extensive air shower can either be directly captured~\cite{Capelle:1998zz} or indirectly observed in forms of radio emission, Cherenkov light, fluorescence and so on~\cite{Jelley1958erenkovRI,1953Nature,Greisen1966,Kampert:2012vi,Schroder:2016hrv,Ginzburg:1945zz,Motloch:2015wca,Revenu:2013lza,deVries:2021BA}.

In those cases discussed, the extensive air shower is considered to be produced from a single tau bang, and the sensitivities to the cosmogenic neutrino flux are estimated based on this assumption~\cite{Neronov:2016zou,GRAND:2018iaj,Otte:2018uxj,Otte:2019aaf,ARA:2019wcf,Abarr:2020bjd,Anker:2020lre,RNO-G:2020rmc,IceCube-Gen2:2020qha,POEMMA:2020ykm,POEMMA:2020ykm,Romero-Wolf:2020pzh,Wissel:2020fav,Wissel:2020sec,Hallmann:2021kqk,Ogawa:2021dK,deVries:2021BA}. 
While this is true at the leading order (LO), 
the novel double- and multi-bang events can arise from the higher-order processes in the SM~\cite{Zhou:2019vxt,Zhou:2019frk,Garcia:2020jwr,Zhou:2021xuh,Soto:2021vdc,Seckel:1997kk,Alikhanov:2014uja,Alikhanov:2015kla}, which cannot be neglected providing the future detection potential.
In general, these processes (see Fig.~\ref{fig:feyn}) are represented by
\begin{eqnarray}
\nu^{}_{\ell} + N \to n\,\tau + X \;,
\end{eqnarray}
where $\ell \in \{e,\mu,\tau \}$ stands for the neutrino flavor and $n \geq 2$ is the number of taus produced in this reaction.
Two or more taus then decay subsequently to produce the double or multiple extensive air showers, within a short time window.
Besides the higher-order processes, multiple tau final states are also predicted by the nonperturbative sphaleron process in the SM, when a certain energy threshold is reached.
Moreover, new physics scenarios such as the microscopic black hole can induce similar outgoing states~\cite{Uehara:2001yk,Alvarez-Muniz:2002snq,Dutta:2002ca,Kowalski:2002gb,Jain:2002kz,Stojkovic:2005fx,Illana:2005pu,Anchordoqui:2006fn,Kisselev:2010zz,Arsene:2013nca,Reynoso:2013jya,Mack:2019bps}.
Note that our double-bang event from di-tau is  different from the ``double-bang''  topology at IceCube~\cite{Learned:1994wg,IceCube:2020abv}, for which the first bang is generated from $\nu^{}_{\tau} \to \tau + X$ and the second bang is from $\tau \to {\rm any}$. For tau neutrino telescopes that observe air showers, the scattering vertex $\nu^{}_{\tau} \to \tau + X$ inside matter is unobservable~\cite{GRAND:2018iaj}.

\begin{figure}[t!]
	\begin{center}
		\includegraphics[width=0.48\textwidth]{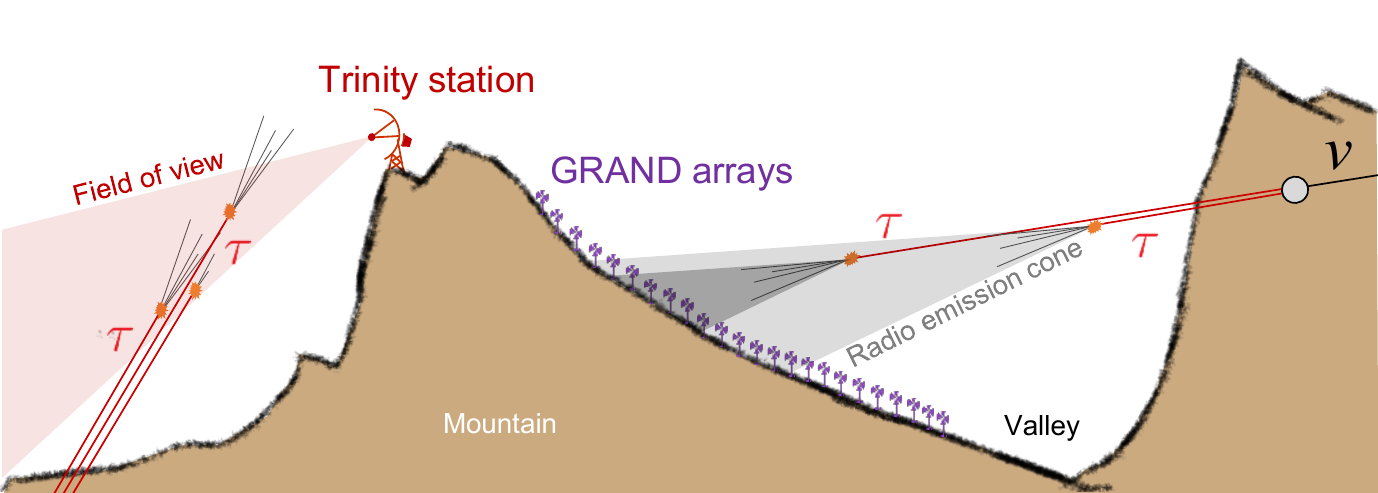}
	\end{center}
	\vspace{-0.3cm}
	\caption{The schematic diagram  (not to scale) of the detection of double and multiple tau bangs, originated from the processes in Fig.~\ref{fig:feyn}. For illustration, two prototypical telescopes, GRAND and Trinity, are taken. }
	\label{fig:schematic}
\end{figure}

In Fig.~\ref{fig:feyn}, we give several representative diagrams relevant for EeV cosmogenic neutrinos.
Diagram (a) is just the leading-order CC scattering, which dominates the neutrino-nucleon cross section above a few TeV.
Diagrams (b), (c1), (c2) and (c3) are capable of producing two and even three taus in a single collision.
Among them, diagram (b) is initiated by a bottom parton in the nucleon~\cite{Soto:2021vdc}, for which a top quark is generated and subsequently decays into a tau.
Whereas diagrams (c1), (c2) and (c3) are initiated by gauge boson contents, i.e., $\gamma$, $Z$ and $W$, inside the nucleus.
In general, the cross section initiated by $\gamma$ is much larger than those by $W$ and $Z$, and can also benefit from the coherent scattering at the nucleus level.
%
The last process, represented by diagram (d), takes place via a nonperturbative sphaleron with the COM energy $\sqrt{s} > E^{}_{\rm sp}\sim 9~{\rm TeV}$. 
%
We will come back to the detailed analyses of those processes in later discussions.


How do the di-tau and multi-tau productions manifest themselves when confronted with the actual experimental configurations?
To see it, we take GRAND~\cite{GRAND:2018iaj} and Trinity~\cite{Otte:2018uxj,Otte:2019aaf,Brown:2021tf,Wang:2021/M} as two representative ground-based telescopes, and discuss their prospects in identifying the double and multiple bangs.

GRAND has a mountain-valley topography, where the radio detection arrays sit on a mountain slope, facing towards another mountain over a valley (see Fig.~\ref{fig:schematic} for a schematic picture).
Radio arrays with $10{\rm k}$ antennas form a projection screen of the area $100 \times 100~{\rm km}^2$, with adjacent antennas separated by a distance of  $1~{\rm km}$.
Providing 20 replicates of such arrays (GRAND200k), an unprecedented sensitivity to the cosmogenic neutrino flux $E^2  \Phi \sim 10^{-10}~{\rm GeV \cdot cm^{-2} \cdot sr^{-1} \cdot s^{-1}}$ can be achieved with a ten-year exposure~\cite{GRAND:2018iaj}.
In contrast, Trinity observes the Cherenkov light emitted from the extensive air shower induced by tau decays.
Instead of containing the whole shower profile like GRAND, Trinity monitors the  activity of only beamed photons but with a wide field of view.
An integrated sensitivity $E^2  \Phi \sim 5.9 \times 10^{-10}~{\rm GeV \cdot cm^{-2} \cdot sr^{-1} \cdot s^{-1}}$ is expected with three stations and ten years of exposure~\cite{Otte:2018uxj,Otte:2019aaf}.


A schematic diagram of the effect of  the double and multiple bangs  is illustrated in Fig.~\ref{fig:schematic}.
Two or three taus are first generated from a single neutrino interaction in matter. After emerged from the mountain or Earth surface, those taus propagate an average distance of $50~{\rm km} \cdot (E^{}_{\tau}/{\rm EeV})$ and then decay in the air.
There is a large randomness in tau decays, which allows the spatial separation of several bangs. Finally, those extensive air showers will be observed within a short coincidence window, and resolved if they are separated enough.
Depending on the actual experimental design, both timing and spatial information can be utilized to resolve the double and multiple bangs.
For the specific GRAND and Trinity setups, some remarks are given as follows.
\begin{itemize}
	\item By design, GRAND antennas will have an excellent time resolution ($\sim 5~{\rm ns}$)~\cite{GRAND:2018iaj}. This is in fact the essential ingredient for GRAND to reconstruct the arrival direction of the extensive air shower, by looking at the timing difference in triggered antennas. As for double bangs, the tau, which decays closer to the radio arrays, will trigger a more concentrated zone. Whereas, the other tau will cover more antennas with the same radio emission cone.
	For some mutual off-axis antennas in coverage of two bangs, there will be characteristically two radio pulses arriving in sequence.
	To set the scale of the typical timing difference, we assume the GRAND screen to sit on a mountain slope inclined by $3^\circ$ and the distance between adjacent antennas to be 1 km. Then, suppose that two taus, emerging from the horizon, decay subsequently around $100~{\rm km}$ away from the arrays and are separated by $5~{\rm km}$. Taking the Cherenkov cone to be $1.5^\circ$, there will be around $400$ antennas triggered in total. For the antenna near the edge of the Cherenkov cone, the timing difference of two bangs is found to be $5.3~{\rm ns}$.
	Given the time resolution power of GRAND, those two pulses should be resolvable. Furthermore, apart from the timing information, the spatial radio footprints of two showers, which overlap with each other, are also promising to resolve with dedicated simulations. This can be seen from GRAND's excellent power in measuring the column depth of the shower maximum $X^{}_{\rm max}$. A target precision of $\sigma (X^{}_{\rm max}) \sim 20~{\rm g \cdot cm^{-2}}$ is already achievable with the radio technique~\cite{Buitink:2014eqa,Corstanje:2021kik}, which roughly corresponds to a resolvable distance of $d = 0.17~{\rm km}$ given the atmosphere density $\rho^{}_{\rm atm} \approx 1.2 \times 10^{-3}~{\rm g \cdot cm^{-3}}$.
	\item For Trinity, multiple separate tau bangs are both resolvable in direction and in timing. Trinity has an excellent field of view around the Earth horizon, and the photon sensitivity is not lost even far outside the typical Cherenkov cone of $1.5^{\circ}$ for the EeV shower. Hence, we do not expect any obvious obstacles for Trinity to identify two or more bangs. The only requirement is from the intrinsic angular resolution $\sim 0.3^{\circ}$ of the camera~\cite{Otte:2018uxj}. 
	This corresponds to a resolvable trajectory distance of $3~{\rm km}$ if the showers are typically $100~{\rm km}$ away from the station and the off-axis angle of showers is $10^{\circ}$. 	The typical probability distributions of the distance between two bangs are given in the Appendix. The probability to have a separation less than $3~{\rm km}$ ($30~{\rm km}$) is  $\sim 10\%$ ($70\%$) for two taus of $0.5~{\rm EeV}$.
    The timing information can also be utilized to distinguish such events from the single bang at Trinity. 
	For the above off-axis shower, the timing difference of two bangs separated by $3~{\rm km}$ is found to be $157~{\rm ns}$.
	If the shower is strictly horizontal, i.e., the telescope is on the axis of the shower, we cannot distinguish several tau bangs. However, we note that such events are very rare in the overall event sample, given the large acceptance angle of Trinity~\cite{Otte:2018uxj}.
	The possible reinforcement of trajectory-sensitive fluorescence detectors will further enhance the ability to reconstruct such events. 
\end{itemize}
Note that the  air shower development occurs along a finite length and will partly smooth away the sharp timing and spacing difference of several bangs. In practice, a likelihood analysis may be necessary to single out such events.
In principle, other ground-based tau neutrino telescopes, e.g.,~Ashra-NTA~\cite{Ogawa:2021dK}, BEACON~\cite{Wissel:2020fav,Wissel:2020sec} and TAMBO~\cite{Romero-Wolf:2020pzh}, also have great potentials in identifying such events with similar setups.

There are also space-borne tau neutrino telescopes, such as POEMMA, deploying satellites in orbit with an altitude of $525~{\rm km}$~\cite{POEMMA:2020ykm}. 
POEMMA in Limb mode monitors the shower activity near the Earth horizon.
However, due to the large distance ($\sim 2640~{\rm km}$) from  satellites to the shower in the atmosphere, it will be challenging to clearly distinguish several bangs with the decay length of $50~{\rm km}$ at EeV energies. Hence, we do not consider such telescopes in this work.

We have seen the potential of ground-based telescopes in identifying the double- and multi-bang events. 
We move on to estimating the event rates of those processes  as in Fig.~\ref{fig:feyn}.

\vspace{0.1cm}

\section{Perturbative di-tau and tri-tau productions}
Due to the relatively small flux of cosmogenic neutrinos at high energies, a large detection volume is required for a positive observation. The detection principle makes tau neutrino telescopes very sensitive to cosmogenic neutrinos around EeV energy scales, where a bump of flux is usually expected. However, as a tradeoff of the large detection volume, below $E^{}_{\nu} = \mathcal{O}(10~{\rm PeV})$  the signal becomes too weak to trigger the detection threshold and the sensitivity of most telescopes falls off.
In such a case, the Glashow resonance from the neutrino-electron scattering 
at $E^{}_{\nu} = 6.3~{\rm PeV}$ will be suppressed for the detection at most tau neutrino telescopes; for relevant discussions, see Refs.~\cite{Fargion:1999se,Fargion:2000iz,Huang:2019hgs,Soto:2021vdc}. Thus, we only focus  on the neutrino-nucleus scattering in this work.

\begin{figure}[t!]
	\begin{center}
		\includegraphics[width=0.48\textwidth]{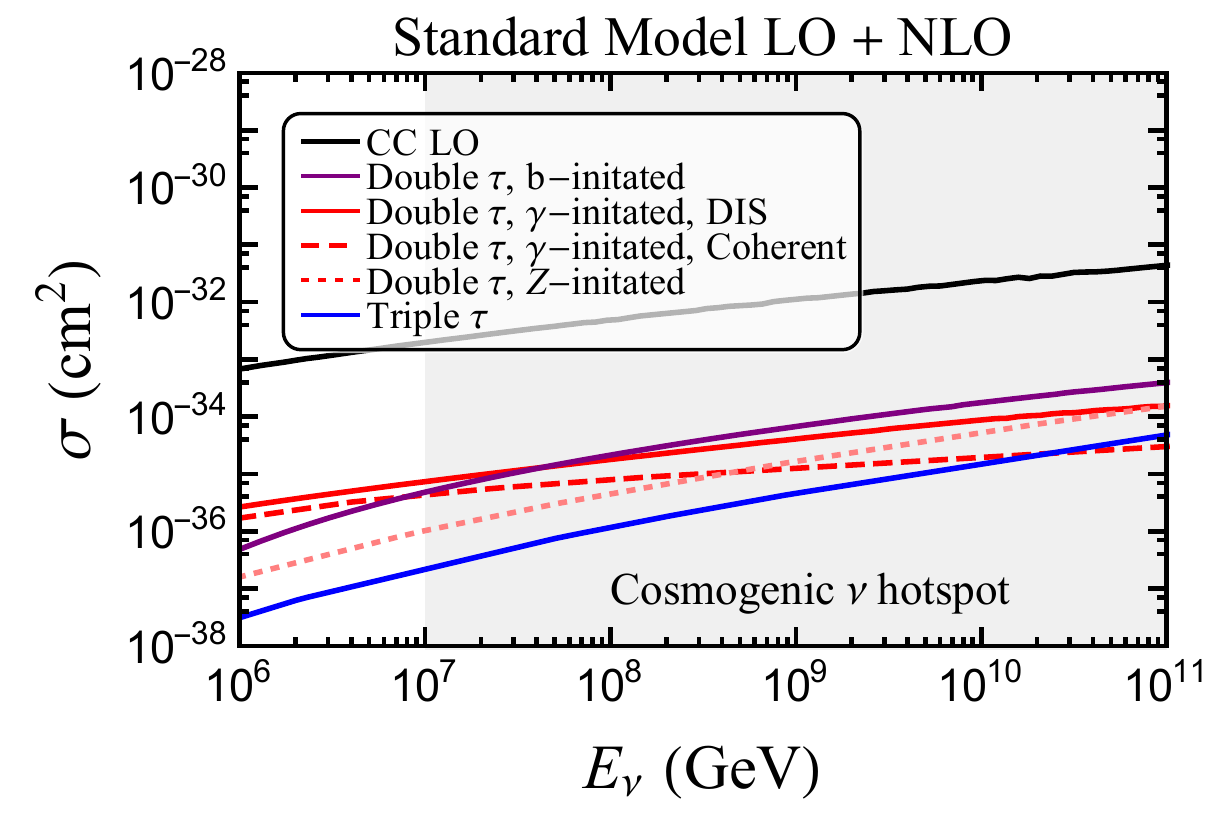}
	\end{center}
	\vspace{-0.3cm}
	\caption{Cross sections of the neutrino-nucleus scattering including: the leading-order CC conversion (black curve), di-tau production initiated by b quark (purple curve), DIS $\gamma$ (red solid curve), coherent $\gamma$ (red dashed curve), or $Z$ (red dotted curve), as well as tri-tau production (blue curve). We take ${}^{16}{\rm O}$ to be the nucleus, and those cross sections are averaged over the nucleon number.}
	\label{fig:xsec}
\end{figure}

\begin{table}[t!]
	\centering
	\begin{tabular}{l |cccc}
		\hline
		Telescopes & 1$\,\tau$ & 2$\,\tau$ & 3$\,\tau$ & {Sphaleron} $n\, \tau$\\
		\hline
		Ashra-NTA~\cite{Ogawa:2021dK} & 19 & 0.2 & 0.007 & 0.7 (0.5)\\
		BEACON~\cite{Wissel:2020fav,Wissel:2020sec} & 137 & 1.6 & 0.062 & 7.1 (5)\\
		GRAND~\cite{GRAND:2018iaj} & 178 & 2.1 & 0.082 & 10 (7)\\
		Trinity~\cite{Otte:2018uxj,Otte:2019aaf} & 16 & 0.2 & 0.006 & 0.6 (0.4)\\
		TAMBO~\cite{Romero-Wolf:2020pzh} & $>7$ & $0.1$ & $0.002$ & $0.11$ (0.08) \\
		[1pt]
		\hline
	\end{tabular} \vspace{0.3cm}
	\caption{ The estimated numbers of single-, double- and triple-bang events from perturbative calculations (first three columns) for several ground-based tau neutrino telescopes with ten years of observation. For illustration, a prediction of the cosmogenic neutrino flux, based on cosmic ray data of Telescope Array~\cite{Anker:2020lre}, is taken as the input to obtain the rough event number. For the sphaleron process, the event number with $n \geq 2$ (or $n \geq 3$ for the number in parentheses) is shown, assuming $\nu^{}_{e}:\nu^{}_{\mu}:\nu^{}_{\tau} = 1:1:1$ and $p^{}_{\rm sp}= 0.003$ with the $s$-wave unitarity bound.
	}
	\label{Tab:number} 
\end{table}

At the leading order, the DIS dominates the neutrino-nucleus cross sections for $E^{}_{\nu}>\mathcal{O}(10~\rm PeV)$. 
Going beyond that, the interest in higher-order processes of the neutrino-matter scattering arises recently~\cite{Zhou:2019vxt,Zhou:2019frk,Garcia:2020jwr,Zhou:2021xuh,Soto:2021vdc,Seckel:1997kk,Alikhanov:2014uja,Alikhanov:2015kla}. 
In particular, the process $\nu^{}_{e,\mu} + N \to \tau + \nu^{}_{\tau}+ X$  was found to contribute a background of a few percent to the detection of cosmogenic tau neutrinos~\cite{Soto:2021vdc}.
However, such a contribution is hard to disentangle from the intrinsic flux uncertainty of cosmogenic neutrinos.

The on-shell $W$ (or $Z$) production becomes kinematically viable when $E^{}_{\nu} > m^{2}_{W}/(2 m^{}_{N}) \approx  3.4~{\rm TeV}$ for neutrino-nucleon scatterings (as illustrated in Fig.~\ref{fig:feyn}), which is different from the suppressed trident production usually discussed~\cite{Altmannshofer:2014pba,Magill:2016hgc,Ge:2017poy,Ballett:2018uuc,Altmannshofer:2019zhy}.
%
The $W$ production in diagrams (c1) and (c2) of Fig.~\ref{fig:feyn} can be initiated by both $\gamma$ and $Z$.
However, because of the collinear enhancement of the massless photon, the major contribution comes from the photon-initiated process at relatively low energy scales.
Diagram (c1) has a tau propagator and is dominant below $E^{}_{\nu} \approx 0.6~{\rm EeV}$, whereas above that diagram (c2) with the $W$-$\gamma$ fusion becomes the dominant channel.
The $Z$ production in diagram (c3) and additional diagrams with the same final state (not shown) is suppressed by the heavy mass of $W$.
We have calculated all possible diagrams and summarize those cross sections in terms of the incoming neutrino energy in Fig.~\ref{fig:xsec}.
The branching ratios ${\rm Br}({W \to \tau \nu_{\tau}}) \approx 11\%$ and ${\rm Br}({Z \to \tau \tau}) \approx 3.4\%$ have been taken into account for our di-tau and tri-tau productions.
For the DIS, the parton distribution functions (PDFs) from CT18qed set~\cite{Xie:2021equ} have been used.

An estimation of the double- and multi-bang events is possible, considering that  higher-order processes only manifest themselves as perturbations to the neutrino and tau fluxes. 
The event number can be roughly estimated by comparing the respective cross section $\sigma^{}_{n\tau}$ to the leading-order one $\sigma^{}_{\rm CC}$. 
Due to the tau decay and energy loss processes, the largest length scale $L^{}_{\tau}$ that a tau can travel in  medium is around $50~{\rm km}$ for all possible energies, which is much shorter than that of a neutrino. Hence, tau events registered in the telescope are essentially produced within a thin layer of $50~{\rm km}$  close to the matter surface.
As a result, the event numbers for both the leading-order and higher-order processes roughly follow the relation 
\begin{eqnarray} \label{eq:N1tauntau}
N^{}_{1\tau/n\tau} \propto \Phi^{\rm out}_{\nu} \cdot \sigma_{\rm CC/n\tau} \cdot L^{}_{\tau} \cdot P^{\rm det}_{1\tau/n\tau}\;, 
\end{eqnarray}
where $\Phi^{\rm out}_{\nu}$ is the exiting neutrino flux near the Earth surface, which can be derived by solving the propagation equation inside matter, and $L^{}_{\tau}$ is the aforementioned  largest length scale that a tau can travel in  the Earth medium, maximally around $50~{\rm km}$ and also limited by the intrinsic matter thickness.
Moreover, $P^{\rm det}_{1\tau/n\tau}$ is the probability that the exiting tau can be accepted by the detector, which depends on its energy and emerging angle. In principle, this probability should be determined by Monte-Carlo simulations including the effects of the tau decay, the shower development as well as the detector response. However, we do not attempt to tackle this issue in the present work. Instead, we only check the significance of the double- and multi-bang events with the ready materials.

By canceling  mutual terms for leading- and higher-order processes in Eq.~(\ref{eq:N1tauntau}), we arrive at
\begin{eqnarray} \label{eq:N1vNn}
\frac{N^{}_{n\tau}}{N^{}_{1\tau}} \approx \frac{\sigma^{}_{\rm n\tau} \cdot P^{\rm det}_{n\tau}}{\sigma^{}_{\rm CC} \cdot P^{\rm det}_{1\tau}}\;. 
\end{eqnarray}
For the single tau case, the detection probability reads $P^{\rm det}_{1\tau} = \int \mathrm{d}s \; p^{}_{\rm decay}(E^{}_{\tau}, s)\;  p^{}_{\rm det}(E^{}_{\tau}, \Omega, s)$, where $s$ is the distance traveled by tau before the decay, $\Omega$ is the incoming angle, $p^{}_{\rm decay}$ is the decay probability and $p^{}_{\rm det}$ is the probability to trigger the detector.
The detector response to the single tau $P^{\rm det}_{1\tau}$ is almost saturated for tau energies higher than $\mathcal{O}(1~\rm EeV)$. As an example, see Fig.~24 of Ref.~\cite{GRAND:2018iaj} for GRAND, where the dependence of the effective detection area on $E^{}_{\nu}$ becomes mild around $(1-10)~{\rm EeV}$.
For two or more taus, the detection probability $P^{\rm det}_{n\tau}$ further depends  on the actual efficiency to distinguish such events from single bang, which should not to be a fundamentally difficult issue as we argued previously.
If the efficiency is of order one and all final-state taus are energetic ($>{\rm EeV}$), we should expect $P^{\rm det}_{n\tau} \sim P^{\rm det}_{1\tau}$, which leads to the rough relation 
${N^{}_{1\tau}}/{N^{}_{n\tau}} \sim  \sigma^{}_{\rm CC}/\sigma^{}_{n\tau}$. 
However, we should note that if the tau energy is too low to trigger the detector, this simple relation will overestimate the double- and multiple-bang events.
In such a case, the actual number requires a dedicated simulation of the detector response, which is, however, beyond the scope of the present work.
In the Appendix, we give the probability distributions of the energy of two taus, and the probability that both taus have energies higher than $E^{}_{\nu}/10$ can be $50\%$. 

Now, let us first estimate the single-bang events based on  given differential sensitivities of GRAND~\cite{GRAND:2018iaj} and Trinity~\cite{Otte:2018uxj,Otte:2019aaf}.
The sensitivities to the cosmogenic neutrino flux from the single bang  are already given for various ground-based telescopes. 
To be definite, we briefly explain in the following how the differential sensitivities were obtained and how to use them to get the actual events for a given input flux. For a specific telescope, the all-flavor sensitivity, defined as the $90\%$ confidence level to observe a positive signal within an energy decade, is recast as~\cite{Reno:2019jtr}
\begin{eqnarray} \label{eq:}
F(E^{}_{\nu}) = \frac{2.44 \times N^{}_{\nu}}{\mathrm{log}(10)\, E^{}_{\nu} \, \langle A\Omega\rangle \, T}\;.
\end{eqnarray}
Here, $2.44$ is the required expectation value of event number, $N^{}_{\nu} = 3$ represents three generations of neutrinos, $\log(10)$ sets the energy interval on the logarithmic scale to be a decade, $\langle A\Omega\rangle$ is the effective aperture of the telescope for tau neutrinos, and $T = 10~{\rm yr}$ is the observational time.
Note that the effective aperture can be obtained reversely if the differential sensitivity is given.

For a given input of the cosmogenic neutrino flux $ E^{2}_{\nu} \mathrm{d} \Phi^{}_{\nu}/ \mathrm{d} E^{}_{\nu}$, the  differential event for the single bang can be obtained with
\begin{eqnarray}
\frac{\mathrm{d}N^{}_{1\tau}}{\mathrm{d} \log^{}_{10} E^{}_{\nu}} = 2.44 \, E^{2}_{\nu}\frac{\mathrm{d} \Phi^{}_{\nu}}{ \mathrm{d} E^{}_{\nu}} \frac{1}{F(E^{}_{\nu})}\;.
\end{eqnarray}
When $ E^{2}_{\nu} \mathrm{d} \Phi^{}_{\nu}/ \mathrm{d} E^{}_{\nu} = F(E^{}_{\nu})$, we reproduce the definition of the differential sensitivity, i.e., $2.44$ events per energy decade.
The $n \tau$ event number can be roughly estimated by using the relation in Eq.~(\ref{eq:N1vNn}). Under the optimistic assumption of $P^{\rm det}_{n\tau} \sim P^{\rm det}_{1\tau}$, we have 
\begin{eqnarray}
\frac{\mathrm{d}N^{}_{n\tau}}{\mathrm{d} \log^{}_{10} E^{}_{\nu}} \sim \frac{\mathrm{d}N^{}_{1\tau}}{\mathrm{d} \log^{}_{10} E^{}_{\nu}} \frac{\sigma^{}_{\rm n\tau} (E^{}_{\nu})}{\sigma^{}_{\rm CC}(E^{}_{\nu})}\;.
\end{eqnarray}
The total event number can be obtained by integrating over the energy.

The obtained numbers of single-, double- and triple-bang events are listed in Table~\ref{Tab:number}, for several ground-based telescopes.
Note that for TAMBO~\cite{Romero-Wolf:2020pzh}, the acceptance to the neutrino flux with $E^{}_{\nu} > 0.1~{\rm EeV} $ is not officially given; therefore the numbers shown there should be interpreted as lower bounds.
For demonstration, we take the cosmogenic neutrino flux  from Ref.~\cite{Anker:2020lre}, which was produced based on recent cosmic ray data of Telescope Array~\cite{Bergman:20197}. We have taken the median flux (i.e., the middle of the flux range on the logarithmic scale) in the allowed statistical range, which is well compatible with the existing constraints on the cosmogenic neutrino flux~\cite{IceCube:2018fhm,PierreAuger:2019ens,ANITA:2019wyx}.
Unless very pessimistic flux predictions are chosen, the neutrino events will not be affected by orders of magnitude by adopting some popular flux choices; see, e.g., Ref.~\cite{Valera:2022ylt}.
In general, with an observation of $\mathcal{O}(100)$ standard single-bang events, one would expect $\mathcal{O}(1)$ double-bang events, whereas the triple-bang rate is rather suppressed and negligible in light of the foreseeable programs.

The observation of double-bang events itself is important for our understanding of how neutrinos interact at ultralight-high energy scales.
Moreover, because such events are very distinctive experimentally, the detectable but suppressed rate actually makes them  an excellent probe of  physics unknowns. One of the examples to be discussed is the sphaleron process.



\section{The sphaleron process and nonperturbative production of many}
The baryon and lepton numbers are violated by nonperturbative processes in the SM~\cite{Adler:1969gk,Bell:1969ts,Bardeen:1969md}, when a transition takes place between vacua characterized by different integer Chern-Simons numbers.
Topologically distinct vacua are separated by an energy barrier of $ E^{}_{\rm sp} \sim m^{}_{W}/\alpha^{}_{W}$ with $\alpha^{}_{W} \approx 1/30$, associated with the sphaleron configuration~\cite{Dashen:1974ck,Manton:1983nd,Klinkhamer:1984di}. 
At low energy scales, the transition between different vacua is possible via the quantum tunneling effect (instanton), which is however exponentially suppressed. In the early Universe,
the transition violating $B+L$ number becomes very efficient beyond tunneling~\cite{Linde:1981zj}, when the temperature can overcome the sphaleron energy barrier $E^{}_{\rm sp}$.
This possibility has been thought to be a crucial ingredient for the baryon asymmetry in our Universe~\cite{Kuzmin:1985mm,Fukugita:1986hr,Rubakov:1996vz,Trodden:1998ym,Buchmuller:2004nz,Davidson:2008bu,Morrissey:2012db,Jaeckel:2022osh}.

The sphaleron process might also be possible for two-particle collisions at nearly zero temperature, when the COM energy can overcome the sphaleron barrier.
The formation of coherent sphaleron state from two-particle collisions was thought to be suppressed~\cite{Ringwald:1989ee,Espinosa:1989qn,Porrati:1990rk,Khlebnikov:1990ue,Khoze:1990bm,Mueller:1991fa,Rebbi:1996zx,Bezrukov:2003er,Bezrukov:2003qm,Ringwald:2002sw,Ringwald:2003ns,Khoze:2020paj}. However, recent studies including the sphaleron periodicity (through multiple sphaleron) suggest counter results~\cite{Tye:2015tva,Tye:2017hfv,Qiu:2018wfb}.
While this fundamental issue itself is still under debate, one can instead turn to the high-energy experimental frontiers for the answer~\cite{Ellis:2016ast,Ellis:2016dgb,Brooijmans:2016lfv,Cerdeno:2018dqk,Jho:2018dvt,Anchordoqui:2018ssd,Zhou:2019uzq,Papaefstathiou:2019djz}.
In particular, possible high-multiplicity events at cosmic ray detectors have been studied~\cite{Jho:2018dvt}, with assumed background rejection methods. At EeV energies, the primary sphaleron process can on average produce $\mathcal{O}(400)$ initial pions in the air shower, compared to the CC/NC one with only $\mathcal{O}(100)$ pions~\cite{Jho:2018dvt}.


\begin{figure}[t]
	\begin{center}
		\includegraphics[width=0.48\textwidth]{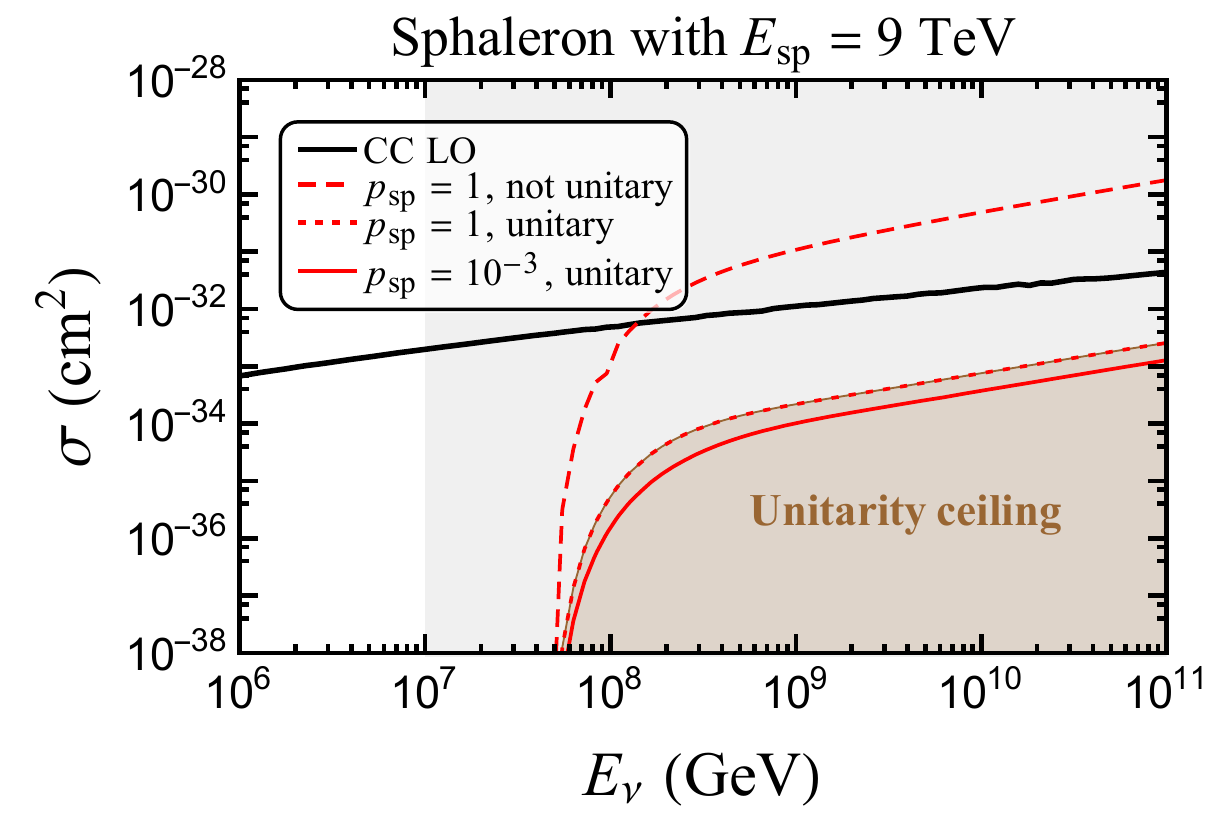}
	\end{center}
	\vspace{-0.3cm}
	\caption{Cross sections of the neutrino-nucleon scattering via the sphaleron process. 
		Without the $s$-wave unitarity requirement, the cross section for $E^{}_{\rm sp}=9~{\rm TeV}$ and $p^{}_{\rm sp}=1$ is given by the red dashed curve. The inclusion of unitarity leads to the red solid ($p^{}_{\rm sp}=10^{-3}$) and red dotted ($p^{}_{\rm sp}=1$) curves, respectively.}
	\label{fig:xsec_sp}
\end{figure}

In our case, the double- and multi-bang topology is very clean in identifying the sphaleron events.
Considering that the COM energy of the neutrino-nucleon collision can be as high as $43~{\rm TeV} \gg E^{}_{\rm sp} \sim 9~{\rm TeV}$ for an EeV incoming neutrino, the detection of cosmogenic neutrinos will be a powerful probe of sphalerons.
The process of our concern reads~\cite{Ringwald:1989ee,Espinosa:1989qn,Gibbs:1994cw}
\begin{eqnarray}
\nu^{}_{\ell} + N \to 2 \overline{l} + 8 \overline{q} + n^{}_{\rm B}\, W/Z/\gamma + X \;,
\end{eqnarray}
where two anti-leptons $\overline{l}$ (can be $\tau$) have flavors  different from the incoming neutrino $\nu^{}_{\ell}$, and $n^{}_{\rm B}$ is the number of gauge bosons produced. 
From the decay of $W/Z$ or top quark, one can have multiple taus in the final state.
The sphaleron reaction features a large multiplicity in the final state, and
the average number of radiated gauge bosons is of the order of $n^{}_{\rm B} \sim 1/\alpha^{}_{W} \approx 30$~\cite{Ringwald:1989ee}.
In our analysis, we do not attempt to get into the sphaleron dynamics of the process, for which a solid theoretical evaluation is lacking thus far. Instead, we can take $n^{}_{\rm B}  \approx 30$ and further assume a flat phase space for the final state, such that the average energy of each outgoing particle  reads $\overline{E}  = E^{}_{\nu} / (10+n^{}_{\rm B}) \approx 0.25~{\rm EeV}\cdot E^{}_{\nu}/(10~{\rm EeV})$. Moreover, the production ratio of $W^+:W^-:Z:\gamma$ is assumed to be $1:1:\cos^2{\theta^{}_{\rm W}}:\sin^2 \theta^{}_{\rm W}$~\cite{Gibbs:1994cw}.

Taking account of the branching ratios of $W\to \tau \nu^{}_{\tau}$ and $Z \to \tau\tau$, the probability to have $n^{}_{} \tau$ with $n \geq 2$ (or $n \geq 3$) from the sphaleron process is high, around $p = 83\%$ (or $61\%$) for  $\nu^{}_{e,\mu}$ and $p = 74\%$  (or $49\%$) for $\nu^{}_{\tau}$ in the initial state. The difference
between the scenarios of $\nu^{}_{e,\mu}$ and $\nu^{}_{\tau}$ stems from the fact that $\nu^{}_{e,\mu}$ has a probability of  producing a primary $\tau$ through the $\Delta L^{}_{\tau} = 1$ sphaleron process while $\nu^{}_{\tau}$ does not.


The parton-level cross section of the sphaleron process can be parameterized as~\cite{Gibbs:1994cw,Ellis:2016ast}
\begin{eqnarray}
\sigma^{}_{0}(\hat{s}) = \frac{p^{}_{\rm sp}}{m^2_{W}} \Theta(\sqrt{\hat{s}} - E^{}_{\rm sp}) \;,
\end{eqnarray}
where $\sqrt{\hat{s}}$ is the total energy in the neutrino-parton COM frame, $p^{}_{\rm sp}$ controls the overall scattering strength, and the Heaviside function is motivated by the consideration of the sphaleron barrier. Since the sphaleron process is approximately pointlike and spherically symmetric, its cross section should be ceiled by the $s$-wave unitarity requirement of the scattering matrix, namely $\sigma^{}_{\rm unitarity} = 16 \pi/\hat{s}$ for $\sqrt{\hat{s}} > E^{}_{\rm sp}$~\cite{Gibbs:1994cw}, which is often ignored. For the phenomenological study, a reasonable choice of the cross section is $\sigma = {\rm min}\{ \sigma^{}_{0}, \sigma^{}_{\rm unitarity}\}$~\cite{Gibbs:1994cw,Papaefstathiou:2019djz}, implying that the higher-order corrections should come to the rescue when the unitarity is saturated. 
The total neutrino-nucleon cross section can then be obtained by convolving the parton-level one with quark PDFs.

An illustration of  cross sections of the sphaleron process is given in Fig.~\ref{fig:xsec_sp} by taking the sphaleron energy scale to be $E^{}_{\rm sp} = 9~{\rm TeV}$.
The most notable message is that the $s$-wave unitarity requirement sets an upper bound to the sphaleron cross section, shown as the brown shaded region. 
A value of $p^{}_{\rm sp} \approx 0.002$ can already saturate the unitarity bound.
Without the unitarity ceiling, the sphaleron process can be even more efficient than the standard CC conversion (not $s$-wave) for $E^{}_{\nu} > 10^8~{\rm GeV}$. Imposing unitarity will considerably reduce the sphaleron cross section, e.g., down to $2\%$ of the CC process at $E^{}_{\nu} = 10^9~{\rm GeV}$.
However, we note that all three neutrino flavors can contribute to the sphaleron process while  only $\nu^{}_{\tau}$ contributes to the CC conversion to tau. 


As shown in Table~\ref{Tab:number}, the combination of several telescopes, capable of collecting $\mathcal{O}(300)$ cosmogenic neutrino events,  can detect $\mathcal{O}(10)$ double and multiple bangs originated from the  sphaleron process even with the unitarity bound. The anticipated signal is beyond the background predicted by our perturbative calculations.
However, we want to emphasize that the results here more or less overestimate the sphaleron event number because the final-state tau energy is degraded by orders of magnitude compared to the initial neutrino. The actual reduction in the event number depends on the shape of the cosmogenic neutrino flux, which can only be answered along with a dedicated detector simulation.









\section{Summary and outlook}
As a flagship option towards the cosmogenic neutrino detection,  tau neutrino telescopes feature an unprecedented detection volume. 
Though  great discovery potentials are provided by such facilities, a broad class of event topologies is lacking  for the particle physics study.

We have explored the potential of the double and multiple bangs, which can be induced by  SM higher-order diagrams as well as nonperturbative processes. 
The topology can also be applied to other interesting scenarios.
It is worthwhile to further investigate the systematic new physics scenarios with  more dedicated experimental simulations in a future work.

\appendix
\section{The probability distributions of energy and separation distance of di-tau}
\noindent
For the convenience of discussion, 
we distinguish different taus in the process of di-tau production: the primary one as $\tau^{}_1$ and the secondary one generated from the $W$ decay as $\tau^{}_2$.

\begin{figure}[t!]
	\begin{center}
		\includegraphics[width=0.4\textwidth]{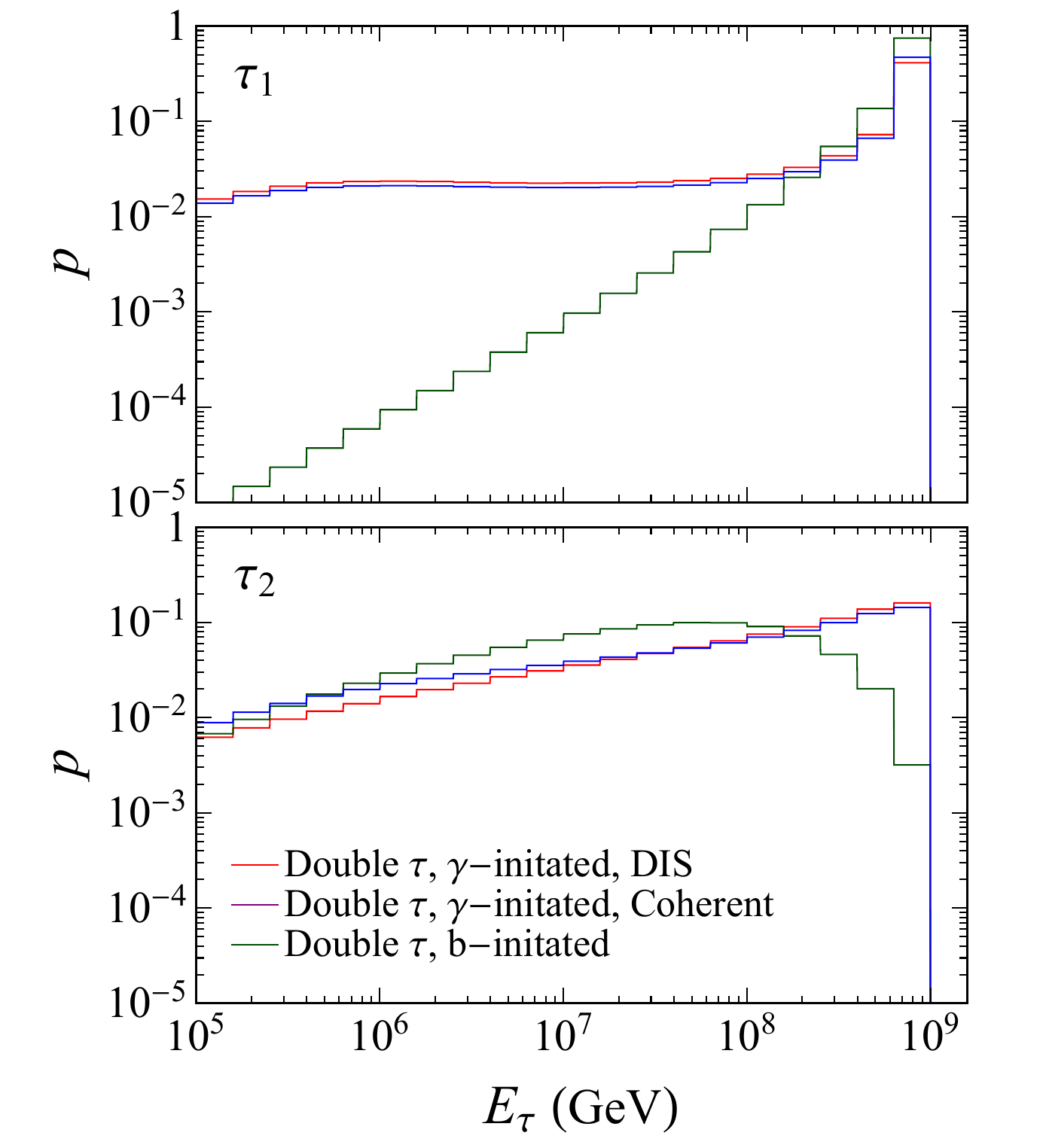}
	\end{center}
	\vspace{-0.3cm}
	\caption{The probability distributions of the energy of two taus produced from different di-tau production processes,  initiated by an EeV incoming neutrino. The primary tau is represented by $\tau^{}_1$ and the secondary one generated from the $W$ decay by $\tau^{}_2$. The probability $p$ is normalized and can be interpreted as the probability to observe tau showers within the energy bin.}
	\label{fig:distri}
\end{figure}

\begin{figure}[t!]
	\begin{center}
		\includegraphics[width=0.4\textwidth]{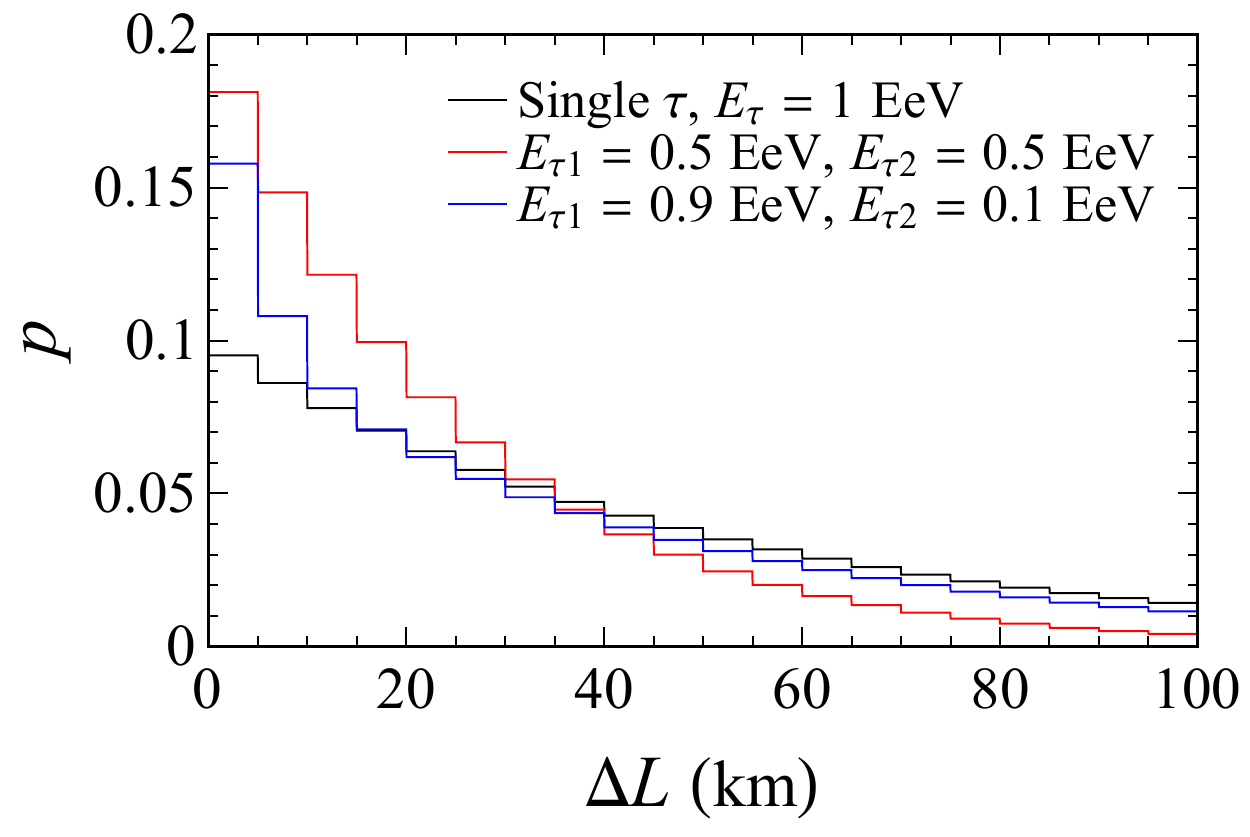}
	\end{center}
	\vspace{-0.3cm}
	\caption{The probability distributions of the distance between two tau bangs (red and blue histograms). For the single tau case, the probability of the decay distance is shown.}
	\label{fig:L_distri}
\end{figure}

Two taus feature distinct energy distributions. 
The energy distributions of two taus can be easily calculated by using the ready differential cross sections of related processes.
For illustration,  we show in Fig.~\ref{fig:distri}  the probability distributions of the energy for two taus initiated by an EeV incoming neutrino. The probability is normalized and assigned to each bin in the histogram plot. 
Different processes are shown separately, including the DIS (in red) and coherent (in purple) processes initiated by $\gamma$ as well as the DIS process initiated by b quark (in green). 	
We notice that the DIS and coherent $\gamma$-initiated processes feature similar distributions. For all processes the phase space is dominated by the small momentum transfer regime (compared to the center-of-mass energy). As a consequence, the primary tau features a higher average energy, while the secondary one from $W$ decays is lower. The integrated probability that the secondary tau has an energy higher than $E^{}_{\nu}/2$ ($E^{}_{\nu}/10$) reads approximately $30\%$ ($50\%$).

Furthermore, the probability distributions of the separation distance between $\tau^{}_1$ and $\tau^{}_2$ decays are shown in  Fig.~\ref{fig:L_distri}. Several benchmark choices are illustrated, including $(E^{}_{\tau 1},E^{}_{\tau 2}) = (0.5,0.5)~{\rm EeV}$ shown in red and $(E^{}_{\tau 1},E^{}_{\tau 2}) = (0.9,0.1)~{\rm EeV}$ shown in blue. For comparison, the decay length of a single tau is illustrated as the black histogram. The probability to have a separation distance larger than $3~{\rm km}$ ($30~{\rm km}$) is as high as $90\%$ ($30\%$) for $E^{}_{\tau 1} = E^{}_{\tau 2} = 0.5~{\rm EeV}$.

\begin{acknowledgements}
The author would like to thank Shun Zhou for very helpful comments on the manuscript. The Feynman diagrams in Fig.~\ref{fig:feyn} are generated with the help of JaxoDraw~\cite{Binosi:2003yf}. The cross sections are calculated in part by using FeynCalc~\cite{Shtabovenko:2020gxv} and CalcHEP~\cite{Belyaev:2012qa} packages. This work is supported by the Alexander von Humboldt Foundation.
\end{acknowledgements}

\bibliographystyle{utcaps_mod}
\bibliography{reference}

\end{document}